\documentclass[11pt,envcountsame]{llncs}
\usepackage{cite}
\usepackage{amsmath}
\usepackage{amssymb}
\usepackage{mathrsfs}
\usepackage{tikz}
\newcommand{\inv}{^{-1}}
\title{The \v{C}ern\'y conjecture for one-cluster automata with prime length cycle}
\author{Benjamin Steinberg\thanks{The author gratefully acknowledges the support of NSERC.  Version of May 11, 2010.}
\institute{School of Mathematics and Statistics\\ Carleton University\\ 1125 Colonel By Drive\\ Ottawa, ON, Canada \\ \email{bsteinbg@math.carleton.ca}}}
\spnewtheorem{Thm}{Theorem}{\bfseries}{\itshape}
\spnewtheorem{Lemma}[Thm]{Lemma}{\bfseries}{\itshape}
\spnewtheorem{Conjecture}[Thm]{Conjecture}{\bfseries}{\itshape}
\spnewtheorem{Rmk}[Thm]{Remark}{\itshape}{\rmfamily}
\spnewtheorem{Prop}[Thm]{Proposition}{\bfseries}{\itshape}
\spnewtheorem{Claim}{Claim}{\itshape}{\rmfamily}
\spnewtheorem{Cor}[Thm]{Corollary}{\bfseries}{\itshape}
\begin{document}
\maketitle

\begin{abstract}
We prove the \v{C}ern\'y conjecture for one-cluster automata with prime length cycle.  Consequences are given for the hybrid Road-coloring-\v{C}ern\'y conjecture for digraphs with a proper cycle of prime length.
\end{abstract}

\section{Introduction}
Let $\mathscr A=(Q,\Sigma)$ be a finite (deterministic) automaton with state set $Q$ and input alphabet $\Sigma$.  A word $w\in \Sigma^*$ is called a \emph{reset word} for $\mathscr A$ if it brings all states to a single state, that is, $|Qw|=1$.  An automaton admitting a reset word is said to be \emph{synchronizing}.   The following conjecture is due to \v{C}ern\'y.

\begin{Conjecture}[\v{C}ern\'y~\cite{cerny}]
An $n$-state synchronizing automaton admits a synchronizing word of length at most $(n-1)^2$.
\end{Conjecture}

The literature on this subject is constantly growing, cf.~\cite{Pincerny,pincernyconjecture,synchgroups,dubuc,cerny,volkovc1,rystsov1,rystsov2,AMSV,trahtman,traht2,volkovc2,Kari,volkovc3,rystcom,rystrank,mycerny,Karicounter,VolkovLata,PerrinBeal,strongtrans,strongtrans2,mortality,beal,Salomcerny,averaging}. The best known upper bound is $(n-1)^3/6$~\cite{twocomb}, whereas it is known that one cannot do better than $(n-1)^2$~\cite{cerny}.

\begin{figure}[hbpt]
	\centering
	\newlength{\nodedist}
	\setlength{\nodedist}{1.5cm}
	
	\begin{tikzpicture}[shorten >=1pt,node distance=\nodedist,auto]
		\tikzstyle{state}=[circle, draw, fill=black!50, inner sep=0pt, minimum width=4pt]
		
		\foreach \X in {0,...,4}
			\node at (\X*72:\nodedist*0.85065)	[state]		(A\X)		{};

		\node	[state]		(B0)		[right of=A0]			{};
		\node	[state]		(B1)		[below right of=B0]		{};
		\node	[state]		(B2)		[below right of=B1]		{};
		\node	[state]		(B3)		[right of=B1]			{};
		\node	[state]		(B4)		[above right of=B0]		{};
		\node	[state]		(B5)		[right of=B4]			{};
		\node	[state]		(B6)		[above right of=B4]		{};
		\node	[state]		(C0)		[above of=A1]			{};
		\node	[state]		(C1)		[above left of=A2]		{};
		\node	[state]		(C2)		[above left of=C1]		{};

		\path [->]
			(A0)		edge		node	[swap]	{$a$}		(A1)
			(A1)		edge		node	[swap]	{$a$}		(A2)
			(A2)		edge		node	[swap]	{$a$}		(A3)
			(A3)		edge		node	[swap]	{$a$}		(A4)
			(A4)		edge		node	[swap]	{$a$}		(A0)
			(B0)		edge		node	[swap]	{$a$}		(A0)
			(B1)		edge		node	[swap]	{$a$}		(B0)
			(B2)		edge		node	[swap]	{$a$}		(B1)
			(B3)		edge		node	[swap]	{$a$}		(B1)
			(B4)		edge		node	[swap]	{$a$}		(B0)
			(B5)		edge		node	[swap]	{$a$}		(B4)
			(B6)		edge		node	[swap]	{$a$}		(B4)
			(C0)		edge		node	[swap]	{$a$}		(A1)
			(C1)		edge		node	[swap]	{$a$}		(A2)
			(C2)		edge		node	[swap]	{$a$}		(C1)
			;

	\end{tikzpicture}
	\caption{$a$-skeleton of a one-cluster automaton with $|Q|=15$ and $|C|=5$.}
	\label{oneclustfig}
\end{figure}
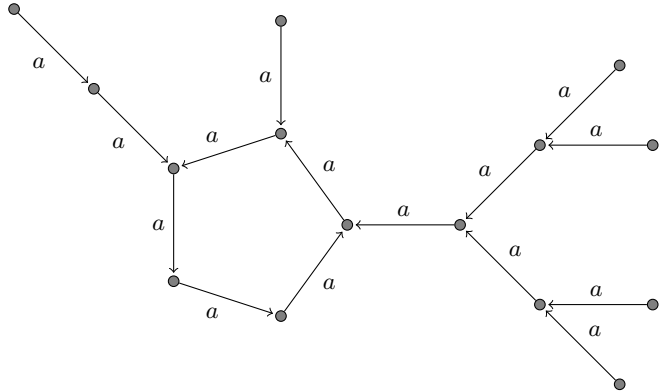

B\'eal and Perrin~\cite{PerrinBeal,bealperrinnew} defined an automaton $\mathscr A=(Q,\Sigma)$ to be a \emph{one-cluster automaton} if there exists $a\in \Sigma$ such that $a$ has only one cycle on $Q$.  More precisely, this means that the graph obtained by considering only the edges labeled by $a$ is connected.  We shall always denote the $a$-cycle by $C$; see Figure~\ref{oneclustfig}.  The level $\ell$ of $\mathscr A$ is the least non-negative integer such that $Qa^{\ell}\subseteq C$.  Of course, $\ell\leq |Q|-|C|$.  If the letter $a$ is not clear from context, then we say that $\mathscr A$ is one-cluster with respect to $a$.

A one-cluster automaton of level $0$, i.e., one in which $Q=C$, is called a \emph{circular automaton}.  The \v{C}ern\'y conjecture was solved by Pin~\cite{Pincerny} for circular automata with a prime number of states and by Dubuc~\cite{dubuc} in the general case (20 years later).   We prove here that the \v{C}ern\'y conjecture is true for one-cluster automata with prime length cycle, generalizing Pin's result.  It is our hope to adapt the techniques of Dubuc~\cite{dubuc} to handle the general case in a later paper.

We also consider the hybrid Road Coloring-\v{C}ern\'y problem, introduced by Volkov.   A strongly connected digraph is said to be \emph{aperiodic} if the greatest common divisor of its cycle lengths is one.  If $\Gamma$ is any strongly connected aperiodic digraph with constant out-degree $|\Sigma|$, then by Trahtman's Road Coloring theorem~\cite{roadcoloring} there is a way to label the edges of $\Gamma$ by $\Sigma$, i.e., to color $\Gamma$, in order to obtain a synchronizing automaton $(Q,\Sigma)$.  The hybrid question is to find the minimum length of a reset word over all possible synchronizing colorings.  As a consequence of our main result, we show that any strongly connected aperiodic digraph with constant out-degree and no multiple edges containing a proper prime length cycle admits a synchronizing coloring with a reset word of length at most $(n-1)^2$ where $n$ is the number of vertices.

\section{Proof of the main result}

Fix for this section a one-cluster synchronizing automaton $\mathscr A=(Q,\Sigma)$ with $n$ states, level $\ell$ and cycle $C$ of prime length $p$ with respect to the letter $a$.  Without loss of generality, we may take $Q=\{1,\ldots,n\}$.  Denote by $\pi\colon \Sigma^*\to M_n(\mathbb Q)$ the matrix representation associated to $\mathscr A$; so \[\pi(w)_{q,r} = \begin{cases} 1 & qw=r\\ 0 & \text{else.}\end{cases}\]  If $S\subseteq Q$, then $[S]$ denotes the characteristic row vector of $S$.  Denote by $v^T$ the transpose of a row vector $v$. Usually, we omit $\pi$ from the notation and write things like $[S]w$ or $w[S]^T$. Note that $w[S]^T= [Sw\inv]^T$ where $Sw\inv = \{q\in Q\mid qw\in S\}$.  If $S\subseteq Q$, let \[\gamma_S= [S]^T-(|S|/p)[Q]^T\] and observe that
\begin{equation*}
w\gamma_S = [Sw\inv]^T-(|S|/p)[Q]^T
\end{equation*}
 for any $w\in \Sigma^*$.  Also we have
\begin{equation}\label{usefuleq}
[C]w\gamma_S = |C\cap Sw\inv|-|S|
\end{equation}
for $w\in \Sigma^*$.

If $V\subseteq \mathbb Q^n$ is a subspace of column vectors, then $\Sigma^*V$ denotes the smallest subspace containing $V$ and invariant under $\Sigma^*$.  If $V$ is spanned by $v_1,\ldots, v_r$, then $\Sigma^*V$ is spanned by the elements $wv_i$ with $w\in \Sigma^*$. The following lemma is a standard ascending chain argument: see for example~\cite{averaging,mycerny,Kari} for a proof.

\begin{Lemma}\label{chain}
Let $\pi\colon \Sigma^*\to M_n(\mathbb Q)$ be a representation and let $W\subseteq V$ be subspaces of $\mathbb Q^n$ consisting of column vectors such that $\Sigma^*W\nsubseteq V$.  Let $A$ be a spanning set for $W$.  Then there exist $w\in \Sigma^*$ and $a\in A$ such that $wa\notin V$ and $|w|\leq \dim V-\dim W+1$.
\end{Lemma}

The next proposition characterizes the fixed column vectors of $a$.  By an \emph{$a$-path} in $\mathscr A$, we mean a path all of whose edges have label $a$.

\begin{Prop}\label{fixedvectors}
Let $v\in \mathbb Q^n$ be a column vector such that $av=v$.  Then $v=k[Q]^T$ for some $k\in \mathbb Q$.
\end{Prop}
\begin{proof}
Routine computation shows that for any state $q$, one has $(av)_q = v_{qa}$.  Thus if $av=v$, then $v_q=v_{qa}$ for any $q\in Q$.  Hence, by iteration, if there is an $a$-path from $q$ to $q'$, then $v_q=v_{q'}$.  But if $q_0$ belongs to the cycle $C$, then every state has an $a$-path to $q_0$.  Thus, all entries of $v$ are the same rational number $k$, that is, $v=k[Q]^T$.
\end{proof}

Our next lemma is crucial in taking advantage of the prime cycle length.

\begin{Lemma}\label{minpoly}
Suppose $\emptyset\neq S\subsetneq C$ and let $A=\{a^{\ell+j}\gamma_{S}\mid 0\leq j\leq p-1\}$.  Let $W$ be the span of $A$.  Then we have: the subspace $W$ is invariant under $\pi(a)$, the minimal polynomial of $\pi(a)$ on $W$ is the cyclotomic polynomial $1+x+x^2+\cdots+x^{p-1}$ and $\dim W=p-1$.
\end{Lemma}
\begin{proof}
Clearly $a^{\ell}$ and $a^{\ell+p}$ act the same on $Q$ and so $\pi(a)^{\ell}=\pi(a)^{\ell+p}$.  It follows that $W$ is invariant under $\pi(a)$, and $\pi(a)^p$ acts on $W$ as the identity.  Therefore, the minimal polynomial $m(x)$ of $\pi(a)$ on $W$ divides $x^p-1$.  Next observe that $W$ is contained in $[C]^{\perp}$ since the fact that $a$ permutes the states of $C$ implies $[C]a^{\ell+j}\gamma_S=[C]\gamma_S= 0$.  Thus $W$ does not contain any non-zero multiple of $[Q]^T$.  It follows from Proposition~\ref{fixedvectors} that $1$ is not an eigenvalue of the restriction of $\pi(a)$ to $W$.  Since $p$ is prime, the factorization over $\mathbb Q$ of $x^p-1$ into irreducibles is $(x-1)(1+x+\cdots+x^{p-1})$. We conclude that $m(x)=1+x+\cdots+x^{p-1}$.  In particular, $\dim W\geq p-1$.  Also, $m(\pi(a))a^{\ell}\gamma_S=0$ implies that $a^{\ell+p-1}\gamma_S=-\sum_{j=0}^{p-2}a^{\ell+j}\gamma_S$ and so $W$ is spanned by $p-1$ elements.  Thus $\dim W=p-1$.
\end{proof}

We need to bound from below the dimension of another subspace.

\begin{Prop}\label{boundconstant}
Let $W=\mathrm{Span}\{a^{\ell}\gamma_q\mid q\in C\}$.  Then $\dim W\geq p-1$.
\end{Prop}
\begin{proof}
Let $C=\{q_1,\ldots,q_p\}$ and define a vector space morphism $\mathbb Q^n\to \mathbb Q^p$ by $[q_i]^T\mapsto e_i$ and $[q]^T\mapsto 0$ for $q\notin C$, where $e_i$ is the $i^{th}$ standard unit vector.  Then the image of $W$ is the space spanned by the vectors $e_j-(1/p)(e_1+\cdots +e_p)$ with $1\leq j\leq p$.  But these vectors form a basis for the orthogonal complement of $e_1+\cdots+e_p$.  It follows $\dim W\geq p-1$.
\end{proof}

Our last lemma relies on Lemma~\ref{chain}.

\begin{Lemma}\label{expand}
Let $\emptyset\neq S_1,\ldots,S_k\subsetneq C$ and $w_1,\ldots,w_k\in \Sigma^*$. Put $W=\mathrm{Span}\{w_i\gamma_{S_i}\mid 1\leq i\leq k\}$. Suppose that \[\sum_{i=1}^kw_i\gamma_{S_i}=0.\]  Then there exist $w\in \Sigma^*$ and $1\leq j\leq k$ such that $|C\cap S_jw_j\inv w\inv|>|S_j|$ and $|w|\leq n-\dim W$.
\end{Lemma}
\begin{proof}
First we claim that there exist $w\in \Sigma^*$ and $1\leq t\leq k$ with $|w|\leq n-\dim W$ and $|C\cap S_tw_t\inv w\inv|-|S_t|\neq 0$.  By \eqref{usefuleq}, this amounts to finding $w$ of length at most $n-\dim W$ and $1\leq t\leq k$ with $ww_t\gamma_{S_t}\notin [C]^{\perp}$.  In particular, we are done if $W\nsubseteq [C]^{\perp}$.  So assume $W\subseteq [C]^{\perp}$.  In particular, $0=[C]w_1\gamma_{S_1} = |C\cap S_1w_1\inv|-|S_1|$ and so $C\cap S_1w_1\inv\neq \emptyset$.  Let $u$ be a reset word.  Since $C=Qa^{\ell}a^*$, we may assume that $u$ resets to a state in $C\cap S_1w_1\inv$.  Then \[[C]uw_1\gamma_{S_1} = |C\cap S_1w_1\inv u\inv|-|S_1| = |C|-|S_1|\neq 0.\]  It follows $\Sigma^*W\nsubseteq [C]^{\perp}$ and so by Lemma~\ref{chain} we can find $w\in \Sigma^*$ and $1\leq t\leq k$ with $ww_t\gamma_{S_t}\notin [C]^{\perp}$ and $|w|\leq \dim [C]^{\perp}-\dim W+1=n-\dim W$.

Next observe that
\[\sum_{i=1}^k (|C\cap S_iw_i\inv w\inv|-|S_i|)  = \sum_{i=1}^k[C]ww_i\gamma_{S_i} = [C]w\sum_{i=1}^kw_i\gamma_{S_i}=0.\]  Since the term $|C\cap S_tw_t\inv w\inv|-|S_t|\neq 0$, there must exist $1\leq j\leq k$ such that $|C\cap S_jw_j\inv w\inv|-|S_j|>0$, that is, $|C\cap S_jw_j\inv w\inv|>|S_j|$.  This completes the proof.
\end{proof}

Our final proposition before proving the main result is a simple computation with derivatives.

\begin{Prop}\label{calculus}
Let $n>0$ be a fixed integer. Then the function \[f(t)= 3n-3t+1+(t-2)(2n-t)\] is bounded by $(n-1)^2$ on the interval $[1,n-1]$.
\end{Prop}
\begin{proof}
We compute $f'(t) = -3+2n-t-t+2=2n-1-2t$ and so is positive for $t<n-\frac{1}{2}$.  Thus $f$ is increasing on the interval $[1,n-1]$ and hence takes its maximum value at $t=n-1$.  Substituting in, we obtain $f(t)\leq 3n-3(n-1)+1+(n-3)(n+1) = 4+n^2-2n-3 = n^2-2n+1=(n-1)^2$.
\end{proof}

We can now prove the main result.

\begin{Thm}\label{main}
Let $\mathscr A=(Q,A)$ be a synchronizing one-cluster automaton with $n$ states, level $\ell$ and cycle $C$ of prime length $p$.  Then $\mathscr A$ has a reset word of length at most \[n-p+1+2\ell+(p-2)(n+\ell)\] which is bounded above by $(n-1)^2$.
\end{Thm}
\begin{proof}
Assume $\mathscr A$ is one-cluster with respect to the letter $a$.  We maintain the above notation.
The proof rests on two claims.
\begin{Claim}\label{claim1} Let $S\subseteq C$ with $2\leq |S|<p$.  Then there exists a word $w\in \Sigma^*$ such that $|C\cap Sw\inv|>|S|$ and $|w|\leq n+\ell$.
\end{Claim}
\begin{proof}[of claim]  Let $W=\mathrm{Span}\{a^{\ell+j}\gamma_S\mid 0\leq j\leq p-1\}$.  Then $\dim W=p-1$ and the minimal polynomial of $\pi(a)$ on $W$ is $m(x)=1+x+\cdots+x^{p-1}$ by Lemma~\ref{minpoly}. Also, \[\sum_{0\leq j\leq p-1} a^{\ell+j}\gamma_S = m(\pi(a))a^{\ell}\gamma_S =0\] because $a^{\ell}\gamma_S\in W$.  Lemma~\ref{expand} now implies we can find $v$ of length at most $n-\dim W=n-(p-1)$ and $0\leq j\leq p-1$ such that \[|C\cap S(a^{\ell+j})\inv v\inv|>|S|.\]  Taking $w=va^{\ell+j}$ does the job because $|w|\leq n-(p-1)+\ell+p-1\leq n+\ell$.
\end{proof}

Our next claim deals with the case $|S|=1$.

\begin{Claim}\label{claim2}
There exists $q\in C$ and a word $w$ of length at most $n-p+1+\ell$ such that $|C\cap qw\inv|>1$.
\end{Claim}
\begin{proof}[of claim]
Let $W= \mathrm{Span}\{a^{\ell}\gamma_q\mid q\in C\}$.  Then $\dim W\geq p-1$ by Proposition~\ref{boundconstant}. Next observe that
\begin{align*}
\sum_{q\in C}a^{\ell}\gamma_q &= a^{\ell}\sum_{q\in C}([q]^T-(1/p)[Q]^T) = a^{\ell}([C]^T-[Q]^T)\\ &= [C(a^{\ell})\inv]^T -[Q]^T= [Q]^T-[Q]^T=0
\end{align*}
where the penultimate equality uses that $Qa^{\ell}=C$.  Lemma~\ref{expand} now provides $q\in C$ and $u\in \Sigma^*$ with $|C\cap q(a^{\ell})\inv u\inv|>1$ and $|u|\leq n-\dim W\leq n-p+1$.  Taking $w=ua^{\ell}$ proves the claim.
\end{proof}

To complete the proof first observe that, by Claim~\ref{claim2}, we can find a state $q\in C$ and a word $w_0$ of length at most $n-p+1+\ell$ such that $|C\cap qw_0\inv|>1$.  Then applying Claim~\ref{claim1}, we can find a word $w_1$ of length at most $n+\ell$ such that \[|C\cap qw_0\inv w_1\inv|\geq |C\cap (C\cap qw_0\inv)w_1\inv|>|C\cap qw_0\inv|.\]  Continuing in this fashion we can find words $w_1,\ldots, w_k$ of length at most $n+\ell$, where $k\leq p-2$, such that $|C\cap qw_0\inv w_1\inv\cdots w_k\inv|=|C|$, i.e., $C\subseteq q(w_k\cdots w_0)\inv$.  Then $Qa^{\ell}w_k\cdots w_1w_0 = Cw_k\cdots w_0 = \{q\}$ and so we have found a reset word $w=a^{\ell}w_k\cdots w_1w_0$ of length at most \[n-p+1+2\ell+(p-2)(n+\ell).\]

Next, using that $\ell\leq n-p$, we obtain an upper bound on $|w|$ of $3n-3p+1+(p-2)(2n-p)$. If $p=n$, then the upper bound becomes $1+(n-2)n=(n-1)^2$.  If $p\leq n-1$, Proposition~\ref{calculus} yields $|w|\leq (n-1)^2$, as required.
\end{proof}

It was observed in~\cite{strongtrans2} that a bound on synchronizing one-cluster automata with prime length cycle leads to bounds for the hybrid Road Coloring-\v{C}ern\'y conjecture.

\begin{Cor}
Let $\Gamma$ be a strongly connected aperiodic digraph with constant out-degree, $n$ vertices and no multiple edges.  Suppose moreover that $\Gamma$ contains a cycle of prime length $p<n$.  Then $\Gamma$ admits a synchronizing word of length at most $3n-3p+1+(p-2)(2n-p)\leq (n-1)^2$.
\end{Cor}
\begin{proof}
Let $C$ be the cycle of length $p$.
A result of O'Brien~\cite{O'Brien,strongtrans2} implies that $\Gamma$ admits a synchronizing coloring that turns it into a one-cluster automaton with $C$ as the cycle.  The result now follows from Theorem~\ref{main}.
\end{proof}

\begin{Rmk}
Note that Lemma~\ref{boundconstant} does not rely on the cycle having prime length.  Using the proof scheme of Theorem~\ref{main} and the ideas of~\cite{averaging}, one can show that if $\mathscr A$ is a one-cluster automaton with $n$ states, level $\ell$ and cycle length $n$, then there is an upper bound of $n-m+2+2\ell+ (m-2)(2n-3)$ on the length of a reset word.  Using that $\ell\leq n-m$ yields an upper bound of $3n-3m +2+(m-2)(2n-3)=m(2n-6)-n+8$.  Assuming $m<n$ (since the case $m=n$ is handled by~\cite{dubuc}), gives an upper bound of $(n-1)(2n-6)-n+8= 2n^2-9n+14$.
\end{Rmk}

\section*{Acknowledgments}
Steffen Kopecki produced the diagram in Figure~\ref{oneclustfig}.

\def\malce{\mathbin{\hbox{$\bigcirc$\rlap{\kern-7.75pt\raise0,50pt\hbox{${\tt
  m}$}}}}}\def\cprime{$'$} \def\cprime{$'$} \def\cprime{$'$} \def\cprime{$'$}
  \def\cprime{$'$} \def\cprime{$'$} \def\cprime{$'$} \def\cprime{$'$}
  \def\cprime{$'$}

\end{document}